\documentclass[pre,showpacs,amsmath,amssymb]{revtex4}

\usepackage{graphicx}
\usepackage{dcolumn}
\usepackage{bm}

\newcommand{\eg}{e.g., }
\newcommand{\Fb}{{F_{\rm b}}}
\newcommand{\ie}{i.e., }
\newcommand{\jss}{{j_{\rm ss}}}
\newcommand{\kT}{{k_{\rm B}T}}
\newcommand{\mmax}{{m^{\rm max}}}
\newcommand{\mmin}{{m^{\rm min}}}
\newcommand{\pd}{\partial}
\newcommand{\phimin}{{\phi_1^{\rm min}}}
\newcommand{\taum}{{\tau_{\rm m}}}

\begin{document}

\title{Premicellar aggregation of amphiphilic molecules:\\ 
Aggregate lifetime and polydispersity}

\author{Radina Hadgiivanova}

\author{Haim Diamant}\email{hdiamant@tau.ac.il}

\affiliation{School of Chemistry, 
Raymond \& Beverly Sackler Faculty of Exact Sciences,
Tel Aviv University, Tel Aviv 69978, Israel}

\date{January 29, 2009}

\begin{abstract}
  A recently introduced thermodynamic model of amphiphilic molecules
  in solution has yielded, under certain realistic conditions, a
  significant presence of metastable aggregates well below the
  critical micelle concentration --- a phenomenon that has been
  reported also experimentally.  The theory is extended in two
  directions pertaining to the experimental and technological
  relevance of such premicellar aggregates. (a) Combining the
  thermodynamic model with reaction rate theory, we calculate the
  lifetime of the metastable aggregates.  (b) Aggregation number
  fluctuations are examined. We demonstrate that, over most of the
  metastable concentration range, the premicellar aggregates should
  have macroscopic lifetimes and small polydispersity.
\end{abstract}



\maketitle

\section{Introduction}
\label{Introduction}

The natural and technological applications of self-assembled
amphiphilic structures (micelles) in aqueous solution are vast
\cite{Wennerstrom,Israelachvili}. According to the common view, 
supported by numerous macroscopic experiments (\eg conductivity and
surface-tension measurements) and widely accepted theories
\cite{Israelachvili,IMN}, amphiphilic molecules form aggregates above
a well defined critical micelle concentration (cmc). During the years,
however, there have been several experimental indications
\cite{Lindman,Estelrich,Zettl,Cui,Estelrich2}, as well as theoretical ones
\cite{Szleifer97}, for the appearance of aggregates at concentrations
well below the cmc --- a phenomenon referred to as premicellar
aggregation. In particular, a fluorescence correlation spectroscopy
experiment \cite{Zettl} seems to have provided direct observation of
premicellar aggregates at concentrations four times lower than the
macroscopically determined cmc.

Recently we have presented a two-state (monomer--aggregate)
thermodynamic model for amphiphilic aggregation which, alongside its
simplicity, allows the study of metastable micelles of variable size
\cite{Radina}. The analysis has yielded a sequence of three well
separated concentrations: $c_1$, where a metastable aggregated state
appears but is not significantly occupied; $c_2$, above which an
appreciable amount of metastable aggregates forms; and $c_3$, where
the aggregated state becomes stable. The cmc as commonly measured in
macroscopic experiments has been shown to correspond to $c_3$. Thus,
appreciable premicellar aggregation may occur in the concentration
range between $c_2$ and $c_3$. We have shown that, so long as the
micelles are not too large, the extent of premicellar aggregation is
much larger than what would be expected from mere finite-size effects.
This somewhat surprising effect stems from the variability of the
``excited state'', \ie the freedom of the micelles to select their
sizes, and the small free-energy difference between the pure monomeric
state and the metastable one, which contains mostly monomers and a low
concentration of aggregates. In addition, we have found that the
premicellar regime is characterized by a weak concentration dependence
of micelle size. Thus, the premicellar aggregates have roughly
the same size as those observed above the cmc, in agreement with the
experiment of Ref.\ \cite{Zettl}.

The analysis in Ref.\ \cite{Radina}, which is purely thermodynamic,
does not fully account for possible kinetic barriers for premicellar
aggregation. It has been assumed, and will be assumed below, that the
solution has fully equilibrated. (In cases where high nucleation
barriers exist, they might be overcome in practice, \eg through
heterogeneous nucleation.)  Two key issues remain open,
however. First, while the metastable premicellar state may be
appreciably occupied at equilibrium, the aggregates might be
short-lived.  Second, although the mean size of the premicellar
aggregates is similar to that of the micelles above the cmc, the size
distribution in the former case might be much broader. Evidently,
these issues of lifetime and polydispersity could jeopardize the
experimental and technological relevance of premicellar aggregation.

Micellization dynamics (above the cmc) were thoroughly studied in
previous works, both experimentally (see Ref.\ \cite{Zana} and
references therein) and theoretically (see, \eg Refs.\ 
\cite{Aniansson1,Aniansson2,Aniansson3,Aniansson4,Kahlweit,Gottberg,CohenStuart,Kuni,Semenov}).
Two disparate time scales are involved in the dynamics, corresponding
to the exchange of individual monomers between the micelle and the
solution and the much slower process of micelle formation and
breakup. Being interested in aggregate stability, we focus here on the
latter.  We use the free energy landscape, as obtained from the
thermodynamic model
\cite{Radina}, within Kramers' rate theory \cite{Kramers,Hangii} to
study the lifetime of premicellar aggregates. The second extension of
the theory is an examination of aggregate size fluctuations in the
premicellar regime.

In Sec.\ \ref{Model} we briefly review the thermodynamic model of
Ref.\ \cite{Radina} and then extend it to study lifetime and
polydispersity. Representative numerical results are presented in
Sec.\ \ref{Results}. In Sec.\ \ref{Discussion} we discuss the results
and their implications for the observation of premicellar aggregates.

\section{Model}
\label{Model}

\subsection{Free energy}
\label{Energy}

Our starting point is the two-state thermodynamic model of
micellization presented in detail in Ref.\ \cite{Radina}. The
solution, containing a volume fraction $\phi$ of amphiphiles, is
assumed to consist of two species: monomers with volume fraction
$\phi_1$, and aggregates of $m$ molecules with volume fraction
$(\phi-\phi_1)$. The volume fraction of water is $(1-\phi)$. Both $\phi_1$
and $m$ are treated as degrees of freedom, \ie the system can select
the number of aggregates as well as their size, while the total volume
fraction $\phi$ is the control parameter. The two-state approximation
restricts the validity of the entire approach to compact (spherical)
micelles, whose size distribution is relatively narrow
\cite{Israelachvili}. The free energy of the solution contains a
mixing-entropy contribution and an interaction term. The former is
calculated using a coarse-grained (Flory-Huggins) lattice scheme,
where a water molecule occupies a single lattice site (of volume
$a^3$), and each amphiphile occupies $n$ sites.  The latter term,
containing all other contributions to the free energy of transfer of a
monomer from the solution to an aggregate of size $m$, is represented
by a single phenomenological function, $u(m)$. The resulting
free-energy density (per lattice site) is
\begin{equation}
F_{\rm s}(\phi_1,m,\phi) = \frac{\phi_1}{n}\ln\phi_1
+ \frac{\phi-\phi_1}{nm}\left[\ln(\phi-\phi_1)-mu(m)\right]
+(1-\phi)\ln(1-\phi).
\label{Fs}
\end{equation}
(All energies in this paper are expressed in units of the thermal
energy $\kT$.)  The specific form of $u(m)$ is not crucial for
the analysis; it should merely have a maximum at a finite value of $m$
to ensure the formation of finite aggregates (rather than a
macroscopic phase) upon increasing $\phi$. For the sake of numerical
examples we shall use the following function \cite{Chandler,Radina}:
\begin{equation}
  u(m) = u_0 - \sigma m^{-1/3} - \kappa m^{2/3}.
\label{u}
\end{equation}
(The physical origins of the terms appearing in Eq.\ (\ref{u}), as
well as the rather limited range of relevant values for the parameters
$u_0$, $\sigma$, and $\kappa$, are discussed in Ref.\ \cite{Radina}.)

Equation (\ref{Fs}) defines a free-energy landscape over a
two-dimensional space of macrostates $(\phi_1,m)$. Along the $\phi_1$
axis $F_{\rm s}$ is always convex, \ie it has a single minimum at
$\phimin(m,\phi)$ for all values of $m$ and for any $\phi$
\cite{Radina}.  Along the $m$ axis the free energy becomes nonconvex
above a certain volume fraction, $\phi>c_1$, with two minima at
$m=1$ and $\mmin$, and a maximum in-between, at $m=\mmax$. In the
premicellar regime of interest, $c_1<c_2<\phi<c_3$, the free energy
has a global minimum still at the pure monomeric state,
$[\phimin(m=1),m=1]$, as well as a local minimum at the metastable
aggregated state, $[\phimin(\mmin),\mmin]$ (containing mostly monomers and a
a low concentration of aggregates).  The two minima are
separated by the saddle point $[\phimin(\mmax),\mmax]$, which poses a
kinetic barrier for the disintegration of the metastable aggregated
state into the stable monomeric one.

The following analysis relies on two basic assumptions. First, we
assume that overcoming the barrier at the saddle point
$[\phimin(\mmax),\mmax]$ is the rate-limiting process in aggregate
dissociation, whereas diffusion is much faster.  Hence, the dynamics
depend on $m$ alone, advancing at all times $t$ along the path
$[\phimin(m(t)),m(t)]$.  The second assumption arises from the
necessity to relate our coarse-grained model with single-aggregate
properties.  Since the model [\eg Eq.\ (\ref{Fs})] does not explicitly
consider single aggregates but rather macrostates containing both
monomers and aggregates, we shall consider, instead, a fictitious
subsystem, of volume $V_1$, which on average contains a single
aggregate of size $\mmin$. The volume of the aggregate itself is
$na^3\mmin$, and the volume fraction of aggregates is $\phi-\phi_1$.
Hence, the subsystem volume is
\begin{equation}
  V_1(\phi) = \frac{na^3\mmin(\phi)}{\phi-\phimin(\mmin(\phi))}.
\label{V1}
\end{equation}
Since $\phi-\phi_1$ is very small, $V_1$ is far from being
microscopic, and we may apply our coarse-grained description to the
subsystem, writing its free energy as
\begin{equation}
  F(\phi_1,m,\phi) = \frac{V_1(\phi)}{a^3} F_{\rm s},
\label{F}
\end{equation}
where $F_{\rm s}$ is given by Eq.\ (\ref{Fs}). Thus, the dissociation
of a single premicellar aggregate is treated as the transition of a
mesoscopic subsystem from a metastable state, containing monomers and
(on average) one aggregate, to the stable, purely monomeric state.
For brevity the free energy of the subsystem along the dissociation
path $[\phimin(m(t)),m(t)]$ is hereafter referred to as $F(m)$.

\subsection{Aggregate lifetime}
\label{lifetime}

We follow the lines of Kramers' theory \cite{Kramers,Hangii} while
adapting it to the case of premicellar aggregates. The main
assumptions of this approach are as follows. (i) The energy barrier
between the two states is sufficiently high, leading to separation of
time scales between the fast monomer exchange process and the much
slower aggregate dissociation. (ii) The free energy of the final
(monomeric) state is much lower than that of the initial (aggregated)
one, ensuring a practically unidirectional probability current from
the aggregated to the monomeric state. The first assumption breaks
down when $\phi$ is too small, \ie as it gets too close to $c_1$; in
the examples of Sec.\ \ref{Results} it becomes invalid already for
$\phi\simeq c_2$.  The second assumption fails when $\phi$ gets close
to $c_3$. Thus, the following calculation is strictly valid only for
$c_2\ll\phi\ll c_3$.  (The behavior outside this domain of validity
will be commented on separately in Sec.\ \ref{Discussion}.) In
addition, we assume that the aggregation number is large, $m\gg 1$, so
that the discrete changes in $m$ can be replaced to a good
approximation by continuous, infinitesimal ones.

We begin with the master equation for the probability density
function, $f(m,t)$, of finding the subsystem around the state
$[\phimin(m),m]$ at time $t$,
\begin{equation}
\label{Master}
\frac{\pd f(m,t)}{\pd t}=
\int dk W(m-k,k)f(m-k,t) - \int dk W(m,k)f(m,t)dk,
\end{equation}
where $W(m,k)$ is the transition probability per unit time for the
aggregation number to change from $m$ to $m+k$. Assuming that large
jumps in aggregation number are improbable, we expand the first
integral in Eq.\ (\ref{Master}) to second order in small $k$ and get
the Fokker-Planck equation,
\begin{equation}
\label{Fokker}
  \frac{\pd f}{\pd t} = -\frac{\pd j}{\pd m},
  \ \ \ 
  j(m,t) = A(m)f(m,t) - \frac{\pd}{\pd m}[D(m)f(m,t)].
\end{equation}
The first term in the probability current density $j$ describes a
drift along the aggregation-number axis, with velocity $A(m)=\int dk
kW(m,k)$. The second term represents diffusion along that axis, with a
diffusion coefficient
\begin{equation}
\label{D}
  D(m) = \frac{1}{2} \int dk k^2 W(m,k).
\end{equation}

Demanding that $f$ reduce at equilibrium (\ie when $j=0$) to the
Boltzmann distribution, $f_{\rm eq}(m)\sim e^{-F(m)}$, one gets from
Eq.\ (\ref{Fokker}) a generalized Einstein relation between $A$ and
$D$,
\begin{equation}
  A(m) = -D(m) F'(m) + D'(m),
\end{equation}
where a prime denotes a derivative with respect to $m$. Substituting
this relation back in Eq.\ (\ref{Fokker}), we rewrite the probability
current density as
\begin{equation}
\label{j}
  j = -D(m) e^{-F(m)} [f(m,t)e^{F(m)}]'.
\end{equation}
Thanks to the assumed high free-energy barrier, and the resulting
separation of time scales, steady state can be assumed practically
throughout the entire dissociation process. Thus, $\pd f/\pd t=\pd
j/\pd m=0$, \ie $j=\jss$ independent of $m$.  Equation (\ref{j}) can
then be integrated over $m$,
\begin{equation}
\label{jint}
  \left. \jss \int_1^{\mmin}dm \frac{e^F}{D} = - f e^F \right|_1^{\mmin}.
\end{equation}

The second assumption, of a large free-energy difference between the
two states, implies that the right-hand side (rhs) of Eq.\ (\ref{jint}) is
dominated by its value at $\mmin$. In addition, we assume that the
subsystem is still mostly in the aggregated state near $m^{\rm min}$,
at quasi-equilibrium, and, hence, $f(m,t)\sim e^{-F(m)}$. Expanding
about $\mmin$ we obtain for the normalized probability density,
\begin{equation}
\label{prob}
  f(m) \simeq \left[ {F''(\mmin)}/{(2\pi)} \right]^{1/2}
  e^{-\frac{1}{2}F''(\mmin)(m-\mmin)^2}.
\end{equation}
The rhs of Eq.\ (\ref{jint}) is given, therefore, by 
$
  -[F''(\mmin)/(2\pi)]^{1/2} e^{F(\mmin)}
$.

Treating the left-hand side (lhs) of Eq.\ (\ref{jint}) requires an
estimate for the aggregation-number diffusion coefficient, $D(m)$. We
use the definition of this coefficient, Eq. (\ref{D}), together with
Langer's formula for the transition probability \cite{Langer},
\begin{equation}
\label{W}
 W(m,k) \sim \tau_0^{-1} e^{-k^2/(2\Delta)}e^{-\frac{1}{2}[F(m+k)-F(m)]}, 
\end{equation} 
where $\tau_0$ is a molecular time scale, and $\Delta$ is used to
suppress large jumps in the aggregation number. Assuming that jumps
much larger than unity are improbable, we set $\Delta=1$. We then
expand $F(m+k)-F(m)$ in Eq.\ (\ref{W}) to second order in $k$,
normalize the transition probability, and substitute it in Eq.\ 
(\ref{D}) to obtain
\begin{equation}
 D(m) = \frac{1}{2\tau_0} \frac{4+F'^2+2F''}{(2+F'')^2}.
\label{d}
\end{equation}
Analysis of Eqs.\ (\ref{Fs})--(\ref{F}) and (\ref{d}) shows that for
realistic aggregation numbers, $m\gg 1$, one has $|F|\gg|\ln D|$.
Hence, the integral on the lhs of Eq.\ (\ref{jint}) is dominated by a
small region around the maximum of $F$. We expand $F(m)$ about
$\mmax$, integrate, and get for the lhs of Eq.\ (\ref{jint}),
$\jss[2\pi/|F''(\mmax)|]^{1/2}e^{F(\mmax)}/D(\mmax)$.

Substituting all these results in Eq.\ (\ref{jint}), we finally obtain
for the micelle lifetime \cite{ft_curvature},
\begin{equation}
 \taum = |\jss|^{-1} = \frac{4\pi\tau_0} {(1+F''(\mmax)/2)
 \left| F''(\mmin)F''(\mmax) \right|^{1/2}} e^{F_{\rm b}},
\label{taum}
\end{equation}
where $\Fb=F(\mmax)-F(\mmin)$ is the height of the free-energy barrier
between the aggregated and monomeric states.  Equation (\ref{taum}),
combined with Eqs. (\ref{Fs})--(\ref{F}), yields the aggregate
lifetime in the metastable, premicellar regime.

\subsection{Polydispersity}
\label{Polydispersity}

For a given amphiphile volume fraction in the premicellar regime,
$c_2<\phi<c_3$, the aggregation number of the metastable aggregates,
$\mmin(\phi)$, is given by the local minimum of $F_{\rm s}$ of Eq.\ 
(\ref{Fs}) \cite{Radina}.  To examine the polydispersity of the
aggregates we should calculate the fluctuations of $m$ around $\mmin$
for a single aggregate. As explained in Sec.\ \ref{Energy}, within our
coarse-grained framework we calculate, instead, the fluctuations of $m$
in a subsystem of volume $V_1$.  The distribution of $m$ in that
subsystem is given, for $m$ close to $\mmin$, by Eq.\ (\ref{prob}).
Thus, we readily get for the mean-square size fluctuation,
\begin{equation}
 \langle\delta m^2\rangle = 1/F''(\mmin).
\end{equation}
The relative width of the size distribution,
\begin{equation}
 w = \frac{\langle\delta m^2\rangle^{1/2}}{\langle m\rangle} 
 = \frac{1}{\mmin [F''(\mmin)]^{1/2}},
\end{equation}
provides a convenient measure of the polydispersity.

\section{Results}
\label{Results}

We now demonstrate the results of the model in two numerical examples,
representing two amphiphiles of differing hydrophobicity.  The
parameters of the amphiphiles are given in Table~\ref{table1},
amphiphile B being the more hydrophobic of the two. We use here the
same two examples whose equilibrium properties have been thoroughly
analyzed in Ref.\ \cite{Radina}.  Table~\ref{table1} lists for these
examples the volume-fraction bounds of the premicellar regime, $c_2$
and $c_3$, along with the aggregation numbers at these points, as
obtained from the equilibrium theory.

\begin{table}[tbh]
\caption{Parameters and equilibrium properties of exemplary amphiphiles.
$n$ --- number of groups in hydrocarbon tail;
$u_0$, $\sigma$, $\kappa$ --- parameters of $u(m)$, the free energy of
transfer in units of $\kT$ [Eq.\ (\ref{u})]. $c_2$, $c_3$ ---
volume-fraction bounds of the premicellar regime; 
$\mmin(c_2)$, $\mmin(c_3)$ --- aggregation numbers at these boundaries 
\cite{Radina}.}
\begin{ruledtabular}
\begin{tabular}{ccccccccc}
amphiphile & $n$ & $u_0$ & $\sigma$ & $\kappa$ & $c_2$ & $c_3$ & $\mmin(c_2)$ & $\mmin(c_3)$\\
\hline
A & 13 & 10 & 11 & 0.08 & $8.0\times10^{-4}$ & $2.2\times10^{-3}$ & 53 & 60\\
B & 20 & 14 & 14 & 0.05 & $1.6\times10^{-5}$ & $6.7\times10^{-5}$ & 118 & 128\\
\end{tabular}
\end{ruledtabular}
\label{table1} 
\end{table}

In Table~\ref{table2} we give the values of the free-energy barrier
for aggregate dissociation at the lower and upper bounds of the
premicellar regime, as calculated from Eqs.\ (\ref{Fs})--(\ref{F})
using the parameters of Table~\ref{table1}. At $\phi=c_2$ the barrier
is negligible, of order $\kT$, yet, as $\phi$ increases through the
premicellar regime, it becomes much larger than $\kT$. The resulting
lifetimes, as calculated from Eq.\ (\ref{taum}), are given in
Table~\ref{table2}. As an estimate for the molecular time scale we
have used for both amphiphiles $\tau_0=10$ ns. (This is the diffusion
time of a molecule, having a diffusion coefficient of $10^{-6}$
cm$^2$/s, along a distance of $1$ nm.) Corresponding to the increase
in the free-energy barrier, the aggregate lifetime increases from
milliseconds at the lower bound of the premicellar region to
practically indefinite time. As already noted in Sec.\ \ref{lifetime},
our lifetime analysis is strictly valid only for $c_2\ll\phi\ll c_3$,
and, hence, these values should be regarded merely as rough estimates.

\begin{table}
\caption{Properties of premicellar aggregates.
$\Fb$ --- free-energy barrier for dissociation;
$\taum$ --- aggregate lifetime;
$w$ --- relative width of size distribution;
$c_2$, $c_3$ --- lower and upper bounds of the premicellar regime.
A value of $\tau_0=10$ ns has been used for the molecular time scale.}
\begin{ruledtabular}
\begin{tabular}{ccccccc} 
amphiphile & $\Fb(c_2)$ & $\Fb(c_3)$ & $\taum(c_2)$ & $\taum(c_3)$ & $w(c_2)$ & $w(c_3)$\\ 
& \multicolumn{1}{c}{$\kT$}&\multicolumn{1}{c}{$\kT$} &\multicolumn{1}{c}{s}&\multicolumn{1}{c}{s}\\
\hline A & 1.3 & 30.4  & $2.0\times10^{-4}$ & $1.3\times10^9$ & 0.18 &
0.15\\ B & 0.5 & 112.5 & $4.3\times10^{-2}$ & $1.8\times10^{43}$ & 0.11 &
0.10\\
\end{tabular}
\end{ruledtabular}
\label{table2} 
\end{table}

\begin{figure}[tbh]
\vspace{0.6cm}
\includegraphics[width=3in]{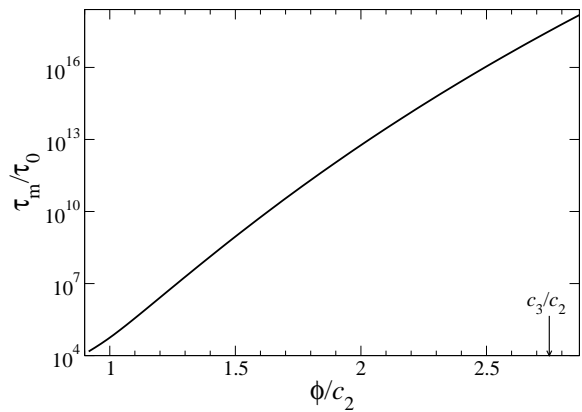}
\caption{Lifetime of premicellar aggregates of amphiphile A as a function
  of amphiphile volume fraction. The lifetime is scaled by the
  molecular time scale $\tau_0$ (of typical order of $10$ ns), and the
  volume fraction by $c_2$, the onset of premicellar aggregation.  The
  volume fraction corresponding to the cmc ($c_3$) is indicated by an
  arrow. Parameters of amphiphile A are given in Table~\ref{table1}.}
\label{fig_lifetime}
\end{figure}

The premicellar aggregate lifetime for amphiphile A, scaled by the
molecular time $\tau_0$, is depicted as a function of amphiphile
volume fraction in Fig.\ \ref{fig_lifetime}. The roughly exponential
increase of lifetime with concentration stems from the exponential
dependence of $\taum$ on the barrier height [Eq.\ (\ref{taum})], which
is the main source of concentration dependence. Two additional
contributions to the dependence of $\taum$ on $\phi$ are included in
the prefactor of Eq.\ (\ref{taum}).  The first,
$(1+F''(\mmax)/2)^{-1}$, comes from the aggregation-number diffusion
coefficient, $D(m)$. This factor is practically
concentration-independent, since in our examples the curvature of the
saddle point is small, $|F''(\mmax)|<0.1$, and thus
$D(\mmax)\simeq(2\tau_0)^{-1}=\mbox{const}$.  The second
pre-exponential factor in Eq.\ (\ref{taum}),
$|F''(\mmin)F''(\mmax)|^{-1/2}$, depends on concentration primarily
through $|F''(\mmax)|$, which is an increasing function of $\phi$.
This factor causes the slightly weaker increase of lifetime with
$\phi$ at small concentration (Fig.\ \ref{fig_lifetime}). 

In Fig.\ \ref{fig_lifetime} we see that in the case of amphiphile A,
assuming $\tau_0\sim 10$ ns, the aggregate lifetime reaches the order
of $1$ s for $\phi\simeq 1.4c_2$, whereas the cmc is at $c_3\simeq
2.75c_2$. Thus, the premicellar aggregates remain stable for a
macroscopic time over a significant part of the premicellar regime. In
the case of amphiphile B we find $\taum\sim 1$ s for $\phi\simeq
1.05c_2$ while $c_3\simeq 4.2c_2$, \ie the aggregates are kinetically
stable over a much larger portion (practically all) of the premicellar
concentration range, as expected for a more hydrophobic
amphiphile.

\begin{figure}[tbh]
\vspace{0.6cm}
\includegraphics[width=3in]{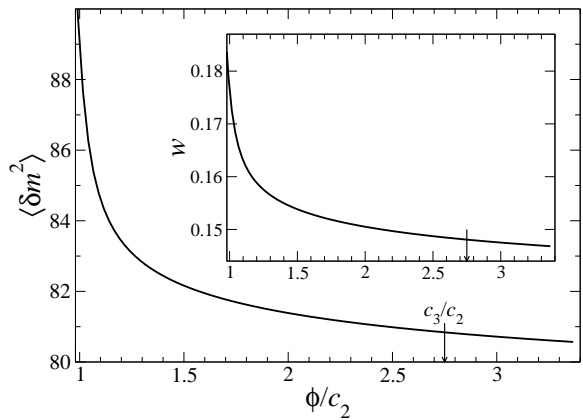}
\caption{Mean-square fluctuation of aggregation number for amphiphile A 
  as a function of amphiphile volume fraction.  The volume fraction is
  scaled by $c_2$, the onset of premicellar aggregation.  The cmc
  ($c_3$) is indicated by an arrow.  The inset shows the relative
  width of the aggregate size distribution, $w=\langle\delta
  m^2\rangle^{1/2}/\langle m\rangle$. Parameters of amphiphile A are
  given in Table~\ref{table1}.}
\label{fig_size}
\end{figure}

Figure \ref{fig_size} shows the mean-square fluctuation of the
aggregation number for amphiphile A as a function of volume
fraction. The corresponding relative width of the aggregate size
distribution is presented in the inset. The polydispersity weakly
decreases with concentration, \ie the premicellar aggregates are
nearly as monodisperse as the micelles above the cmc. In
Table~\ref{table2} we see that the same conclusions hold for
amphiphile B. The small polydispersity (around 10\%), as well as the
slightly increased value for the less hydrophobic amphiphile (A), are
in agreement with the well known trends for spherical micelles above
the cmc, as established experimentally \cite{Zana} and theoretically
\cite{Israelachvili}.

\section{Discussion}
\label{Discussion}

It has been shown in Sec.\ \ref{Results} that considerations of
aggregate lifetime can reduce the concentration range in which
premicellar aggregates may be experimentally observable and
technologically relevant, compared to the range determined from
equilibrium considerations alone \cite{Radina}. In other words, the
apparent concentration, above which an appreciable amount of
metastable micelles appears, may be higher than $c_2$. We have
demonstrated, nonetheless, that kinetic stability (\ie macroscopic
lifetime) still exists in most of the premicellar region. The more
hydrophobic the surfactant, the wider the range of stability is. These
conclusions are in line with results presented in Ref.\
\cite{Semenov}. Although that study does not deal with premicellar
aggregation, it has shown that the dissociation time of micelles
remains very large even below the cmc.

Our Kramers-like approach, as already mentioned in Sec.\ \ref{Model},
relies on two assumptions, which are violated near the edges of the
premicellar region. The first assumption, of a high free-energy
barrier between the metastable and stable states, is valid in almost
the entire region except close to the lower edge, $c_2$, where the
barrier may become of order $\kT$ only. (See Table~\ref{table2}.) The
resulting short lifetimes, though not accurately accounted for by the
theory, are of little interest. The second assumption, of a large
free-energy difference between the two states, holds in nearly the
entire range as well, except very close to the upper edge, $c_3$,
where, by definition, the free-energy difference vanishes.  The
free-energy difference, in units of $\kT$, becomes large quickly as
$\phi$ gets smaller than $c_3$, since the considered mesoscopic
subsystem of volume $V_1$ contains a large number of molecules (mostly
monomers). In addition, correction of the theory near $c_3$ by
considering a probability backflow from the monomeric to the
aggregated state will only increase the stability of the
latter. Therefore, the deficiencies of the theory at the edges of the
premicellar region do not affect our main results.

It should be borne in mind, however, that the stability of premicellar
aggregates imply also that high nucleation barriers may need to be
overcome in order for them to form in the first
place. Correspondingly, the more hydrophobic the surfactant, the
higher these barriers are. (The issue of high nucleation barriers for
micelle formation above the cmc has been underlined also in Ref.\
\cite{Semenov}.) Hence, since both the preceding work \cite{Radina}
and the current one have assumed full equilibration, their
applicability within reasonable time scales might require in practice
either reduction of nucleation barriers through heterogeneous
nucleation or overcoming them by external means (\eg agitation or
sonication).

Finally, we have found narrow size distributions of premicellar
aggregates, \ie micelles below the cmc should be only slightly more
polydisperse than their counterparts above the cmc. (See Fig.\ 
\ref{fig_size}.) This agrees with the monodispersity observed in
experiment \cite{Zettl}.  Thus, polydispersity does not pose a problem
for the applicability of premicellar aggregation.

\begin{acknowledgments}
  RH would like to thank R.\ Metzler and the Technical University of
  Munich for their hospitality. This research was supported in part by
  the Israel Science Foundation (Grant No.\ 588/06).
\end{acknowledgments}


\begin{thebibliography}{99}
\bibitem{ft_curvature} For $F''(\mmax)\leq -2$ this result evidently
  breaks down. In the representative examples of Sec.\ \ref{Results} we
  have $|F''(\mmax)|<0.1$. One can find examples where $|F''|$
  approaches 2, yet this occurs only close to $c_3$, the upper edge of the
  metastable region, where the Kramers-like approach becomes
  inappropriate.  Technically, the divergence can anyway be avoided by
  reducing the value of the parameter $\Delta$ in Eq.\ (\ref{W}).
\end{thebibliography}
\end{document}